\documentclass[
   reprint,
   superscriptaddress,
   amsmath,amssymb,
   aps,
   pra,
   longbibliography,floatfix
]{revtex4-2}

\usepackage{braket}
\usepackage{adjustbox}

\usepackage{graphicx}
\usepackage{amsmath}
\usepackage{amssymb}
\usepackage{float}
\usepackage[dvipsnames]{xcolor}
\usepackage[colorlinks=true,urlcolor=blue,linkcolor=blue,citecolor=blue]{hyperref}
\usepackage[capitalize]{cleveref}
\usepackage{physics}
\usepackage{bm}
\usepackage{dblfloatfix}

\usepackage{tikz}
\usetikzlibrary{quantikz}

\DeclareFontFamily{U}{wncy}{}
\DeclareFontShape{U}{wncy}{m}{n}{<->wncyr10}{}
\DeclareSymbolFont{mcy}{U}{wncy}{m}{n}
\DeclareMathSymbol{\Sh}{\mathord}{mcy}{"58} 

\usepackage{mathtools}

\begin{document}

\title{Optimization of Two-Qubit Gates in Tunable-Coupler Architectures \\Using Single Flux Quantum Control}

\author{Boyan~Torosov}
\affiliation{1QB Information Technologies (1QBit), Vancouver, British Columbia, Canada}

\author{Bohdan~Kulchytskyy}
\affiliation{1QB Information Technologies (1QBit), Vancouver, British Columbia, Canada}

\author{Florian~Hopfmueller}
\affiliation{1QB Information Technologies (1QBit), Vancouver, British Columbia, Canada}

\author{John~Gunderson}
\affiliation{1QB Information Technologies (1QBit), Vancouver, British Columbia, Canada}

\author{Xiangzhou~Kong}
\affiliation{1QB Information Technologies (1QBit), Vancouver, British Columbia, Canada}

\author{Pooya~Ronagh}
\email[Corresponding author: ]{pooya.ronagh@1qbit.com}
\affiliation{1QB Information Technologies (1QBit), Vancouver, British Columbia, Canada}
\affiliation{Institute for Quantum Computing, University of Waterloo, Waterloo, Ontario, Canada}
\affiliation{Department of Physics \& Astronomy, University of Waterloo, Waterloo, Ontario, Canada}
\affiliation{Perimeter Institute for Theoretical Physics, Waterloo, Ontario, Canada}

\begin{abstract}
We present a gradient-based method to construct high-fidelity, two-qubit quantum gates in a system consisting of two transmon qubits coupled via a tunable coupler. In particular, we focus on single flux quantum (SFQ) pulses as a promising and scalable alternative to traditional control schemes that use microwave electronics. We develop a continuous embedding scheme to optimize these discrete pulses, taking advantage of auto-differentiation of our model. This approach allows us to achieve fSim-type gates with average gate fidelities on the order of 99.99\%  and CZ and CNOT gates with fidelities above 99.9\%. Furthermore, we provide an alternative semi-analytical  construction of these gates via an exact decomposition using a pair of fSim gates which leads to the reduction in memory required to store the associated pulse sequences. 

\end{abstract}
\date{\today}

\maketitle

\section{Introduction}

Achieving fault-tolerant quantum computation (FTQC) represents a critical milestone in realizing practical quantum computing, offering the potential to solve complex problems beyond the reach of classical systems. However, FTQC necessitates addressing significant challenges, including efficient input/output interfacing, scalable wiring, and minimizing control overhead. In this context, superconducting single flux quantum (SFQ) technology has emerged as a promising solution, offering fast, energy-efficient digital control and readout capabilities that align well with the stringent requirements of FTQC systems. 

Traditionally, quantum processors with superconducting qubits are controlled via a classical controller, which sends microwave voltage pulses to each qubit to perform a certain operation. This approach suffers from limitations associated with the difficulties in generating and sending the microwave signals from the classical controller at room temperature to the quantum chip in the dilution refrigerator. Furthermore, heat dissipation at millikelvin temperatures is also an issue to consider~\cite{krinner2019engineering}, due to the limited cooling power of the refrigerator, as well as the dissipative nature of certain components, such as attenuators and filters.

One possible solution to these limitations is to use SFQ technology~\cite{lin1995timing, mancini1999phase} and move the controller to the refrigerator. In this case, qubits are controlled via a train of short voltage pulses, each with an area equal to the magnetic flux quantum $\Phi_0=h/2e$. This approach has been proposed as a scalable alternative to microwave control, and has been used to implement single-qubit and two-qubit gates, obtaining fidelities comparable to those attained with microwave control~\cite{leonard2019digital, li2019hardware, mcdermott2014accurate, howe2022digital, liu2023single, liebermann2016optimal, mcdermott2018quantum, dalgaard2020global, jokar2021practical, jokar2022digiq, wang2023single}.

In our study, we present a method to construct high-fidelity two-qubit gates via SFQ control using a tunable-coupler architecture~\cite{Yan2018}, which has become a preferred approach for two-qubit gate implementation~\cite{GoogleSupremacy, google2023suppressing, stehlik2021tunable, sete2024error, sung2021realization}.
We develop two strategies for optimal control using SFQ sequences, either by a direct gradient-based optimization, or by using an analytical decomposition of a CZ or CNOT gate into fSim gates combined with single-qubit gates. This approach differs from earlier SFQ optimization techniques, which primarily rely on discrete optimization methods such as genetic algorithms~\cite{liebermann2016optimal, jokar2021practical}. These algorithms frequently become inefficient due to the large number of variational parameters and result in unstructured solution sequences that lead to large memory requirements.
We target three types of two-qubit gates: an fSim-like gate, a CZ gate, and a CNOT gate. Furthermore, we apply previously developed single-qubit SFQ sequences~\cite{shillito2023compact} in our two-qubit tunable-coupler architecture, which lays the groundwork for a modular and scalable controller. The performance of our sequences is comparable to the state of the art, with average a gate fidelity close to 99.99\% for the fSim gate, and between 99.9\% and 99.99\% for the CZ, CNOT, and single-qubit gates, making it a promising candidate for FTQC. 

This paper is organized as follows. In \Cref{sec:Hamiltonian} we introduce the lumped-element circuit of the tunable-coupler architecture and examine its Hamiltonian. \Cref{sec:simulation} explains how the system is simulated of, namely, how we choose the simulation and logical bases. In \Cref{sec:method}, we describe our optimization procedure, and in \Cref{sec:results}, we present the fidelity values achieved for the two-qubit gates considered.
We summarize the results of our study in \Cref{sec:conclusion}.

\section{Hamiltonian of a Tunable Coupler}\label{sec:Hamiltonian}

\begin{figure}[ht]
\centerline{\includegraphics[width=1.05\columnwidth]{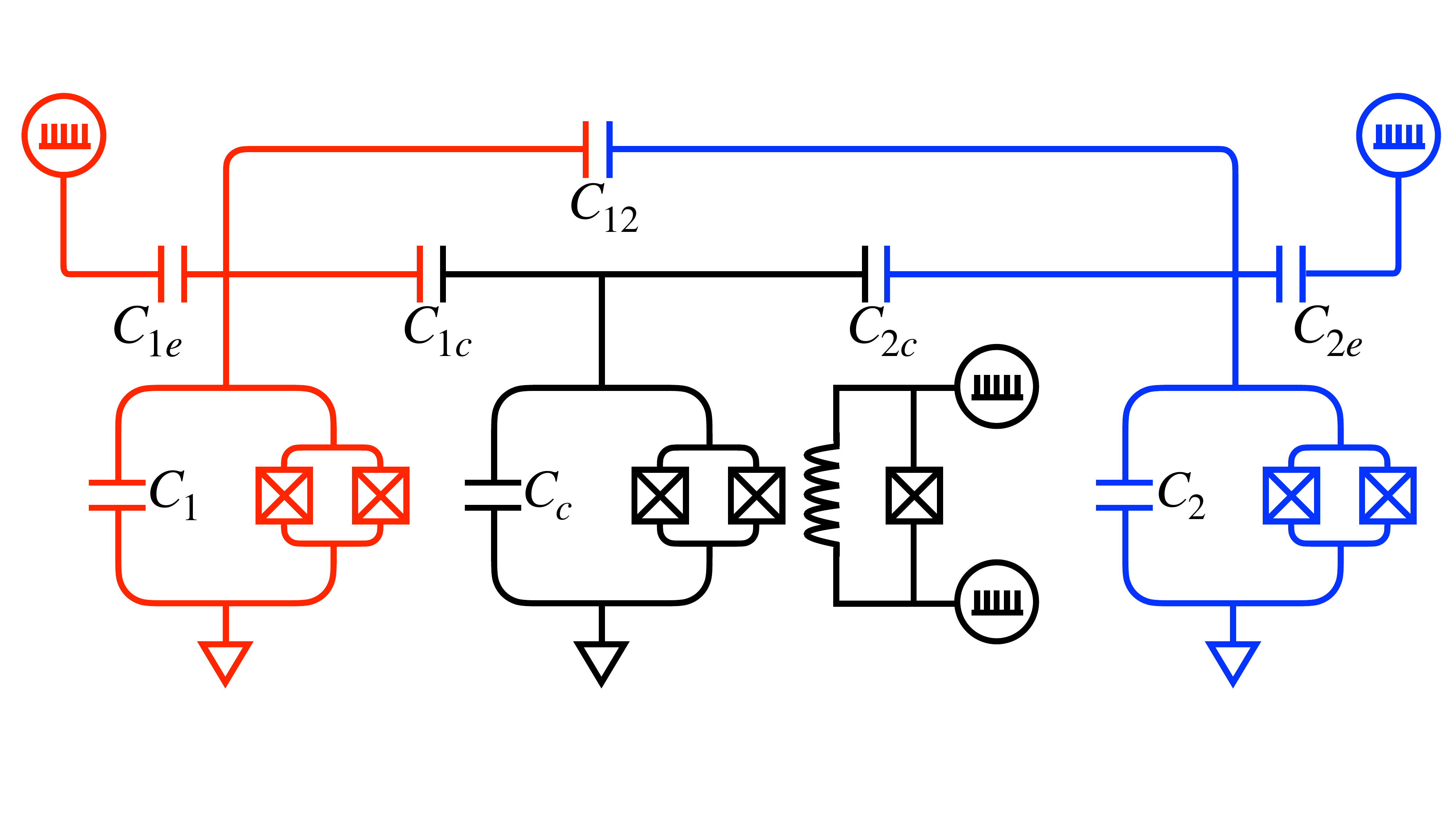}}
\caption{Circuit diagram of a tunable coupler, similar to that of Ref.~\cite{Yan2018}: two transmon qubits (red and blue) are capacitively coupled both directly and also to a coupler qubit (black). The qubits are capacitively driven by sequences of SFQ pulses, which provide single-qubit control, while the coupler is inductively driven using an SFQ circuit to add and remove flux portions into the coupler qubit's loop~\cite{kirichenko2022system}. }
\label{fig:tun_coupler}
\end{figure}

The system that we study consists of a pair of transmon qubits coupled via a tunable coupler~\cite{Yan2018}, which modulates the interaction strength between the qubits. The tunable coupler is in fact also a transmon qubit, and the frequency of all three transmons can be tuned using external fluxes. In addition, the two qubits are capacitively coupled to an SFQ pulse generator, and the coupler qubit is inductively coupled to an SFQ circuit~\cite{kirichenko2022system} capable of adding and removing flux portions to the coupler loop. A schematic representation of the system's circuit is presented in \cref{fig:tun_coupler}. The Hamiltonian of the system can be derived using circuit QED methods~\cite{vool2017introduction}: starting from the classical Lagrangian, we write the classical Hamiltonian using Legendre transformation, and then quantize the obtained classical Hamiltonian. The resulting Hamiltonian $H$ can be split into three parts, that is, single-qubit Hamiltonians $H_k$, a coupling Hamiltonian $H_c$, and a drive Hamiltonian $H_d$, represented as 
\begin{equation}\label{JointHamiltonian}
    H = \sum_{k}H_k + H_c + H_d.
\end{equation}
The single-qubit Hamiltonians are given by
\begin{equation}
    H_k = 4 \hat{n}_k (E_C)_{k,k} \hat{n}_k - E_{J,k} \cos \hat{\phi}_k ,
\end{equation}
where $k=1,c,2$ is an index that runs over the qubits (the first qubit, the coupler qubit, and the second qubit, respectively) $\hat{n}_k$ and $\hat{\phi}_k$ are the reduced charge and flux operators, and $E_{J,k}$ are the Josephson energies that depend on the external fluxes through the loops. The coupling matrix $\mathbf{E}_C$ is proportional to the inverse of the capacitance matrix
\begin{widetext}
    \begin{equation}
    \mathbf{M}_C = 
    \begin{bmatrix}
        C_{1e}+C_1+C_{1c}+C_{12} & -C_{1c} & -C_{12} \\
        -C_{1c}  & C_c+C_{1c}+C_{2c}  & -C_{2c} \\
        -C_{12} & -C_{2c} &  C_{2e}+C_2+C_{2c}+C_{12} 
    \end{bmatrix}
    \end{equation}    
\end{widetext}
whose corresponding capacitances are illustrated in \cref{fig:tun_coupler}.
The second term
\begin{equation}
    H_c = 4 \sum_{k, l\neq k} \hat{n}_k (E_C)_{k,l} \hat{n}_l
\end{equation}
couples the charge operators of the two transmon qubits and the coupler qubit, and the drive term
\begin{equation}
    H_d = -8 \Vec{n}^e \cdot \mathbf{E}_C \cdot \Vec{\hat{n}}
\end{equation}
couples the charge operators $\Vec{\hat{n}}=[\hat{n}_1,\hat{n}_c,\hat{n}_2]^T$ to the external SFQ pulses, represented by $\Vec{n}^e=[n_1^e,0,n_2^e]^T$.
The latter can be approximately described as sequences of delta function-shaped voltage spikes~\cite{mcdermott2014accurate}. 
We note that the drive term describes only the single-qubit drive, while the coupler control is implicitly described by the single-qubit Hamiltonian whose Josephson-energy terms depend on the value of the external flux. This flux control is performed in an SFQ fashion also, by adding and removing flux portions to the coupler-qubit's loop via a sequence of discrete steps.

\section{Simulation of the System Hamiltonian}\label{sec:simulation}

We now describe how the numerical simulation of the studied system is performed, which is then used to conduct SFQ-control optimization for high-fidelity, two-qubit gates. 

\subsection{Simulation Basis}

To simulate the Hamiltonian of the system, we need to introduce a truncated basis. There are different ways to do this, such as using either the Fock, charge, or flux basis. 
For our simulations, we have chosen the charge basis
\begin{equation}
    \hat{n} = \sum_{n=-\infty}^{\infty} n \ketbra{n}{n}
\end{equation}
as a simulation basis, which has the benefit of treating the periodicity of the variable $\hat{\phi}$ correctly. Using the commutation relation between the charge and flux variables, one can derive the explicit expressions for the cosine and sine of the flux variable in the charge basis:
\begin{subequations}
    \begin{align}
        \cos\hat{\phi} &= \frac{1}{2} \sum_{n=-\infty}^{\infty} \left(\ketbra{n}{n+1} + \ketbra{n+1}{n}\right), \\
        \sin\hat{\phi} &= \frac{1}{2} \sum_{n=-\infty}^{\infty} \left(\ketbra{n}{n+1} - \ketbra{n+1}{n}\right).
    \end{align}
\end{subequations}
Next, we apply the following procedure:
\begin{itemize}
  \item Express the single-qubit terms $H_k$ in this basis, truncating at some value of $n$, e.g., $n=\pm 50$.
  \item Diagonalize numerically, setting the phase of the eigenvectors $\ket{\psi_i}$ such that $\bra{\psi_i}\hat{n}\ket{\psi_{i+1}}$ is negative imaginary (similar to $\hat{n}$ in the Fock basis).
  \item Express $\hat{n}$ and $H_k$ in this basis.
  \item Truncate to the first few eigen levels. In our simulations, we truncate to five levels.
\end{itemize}
To deal with time-dependent flux, we note that the coupler's control is implemented via a finite, discrete set of flux values. Hence, we can diagonalize the Hamiltonian at each flux value and calculate the basis-change unitaries that translate between eigenbases at the different values.

After building the simulation basis out of single-qubit Hamiltonian bases, and prior to running the optimizations, we must adjust the values of the external fluxes of the frequency-tunable transmons. During the execution of quantum circuits, these are the flux values at which idling will occur. We will call these external fluxes \emph{idling fluxes} or \emph{off-point fluxes}. Therefore, our goal is to find the flux values that cause the two qubits to have the same target frequency, and for their effective coupling to become zero.
This calibration is obtained through numerical optimization, which minimizes the difference between the first and second excited energies of the joint Hamiltonian, and sets them equal to a given target value. In our optimization, this target value is $\omega_1=\omega_2=2\pi\times 5$~GHz.

Finally, we note that the simulated Hamiltonian naturally depends on the specific values of the capacitances and Josephson junction currents of the circuit depicted in \cref{fig:tun_coupler}. For our simulations, the values of the capacitances are
\begin{align*}
    & C_1=C_2 = 70~\mathrm{fF},\\
    & C_c = 60~\mathrm{fF},\\
    & C_{12} = 0.25~\mathrm{fF},\\
    & C_{1c} = C_{2c} = 2~\mathrm{fF},
\end{align*}
and the values of the Josephson currents are
\begin{align*}
    & J_{1L} = J_{2L} = 7~\mathrm{nA},\\
    & J_{1R} = J_{2R} = 21~\mathrm{nA},\\
    & J_{cL} = 18~\mathrm{nA},\\
    & J_{cR} = 36~\mathrm{nA},
\end{align*}
where $J_{kL}$ and $J_{kR}$ are the critical currents of the left and right Josephson junctions of qubit $k$, with $k=1,c,2$. In addition, the optimized off-point and on-point flux values are
\begin{align*}
    & \Phi_{\textrm{off}} \approx [0.130, 0.352, 0.130]~\Phi_0,\\
    & \Phi_{\textrm{on}} \approx [0.130, 0.376, 0.130]~\Phi_0,
\end{align*}
where the three vector components correspond to qubit~1, the coupler, and qubit~2, respectively.

\subsection{Logical Basis}

As explained in the previous section, the simulation basis is built from single-qubit Hamiltonian eigenstates. We note that the eigenstates of the total Hamiltonian~\eqref{JointHamiltonian}, even at the off-point and without drives, are not product states. Therefore, the next step is to define our computational basis, which determines the logical states. We do this such that idling errors are minimized. Namely, we define the logical states as eigenstates of the joint Hamiltonian at idling fluxes and zero drive. This ensures that there are no state transitions or leakage during idling, though some conditional phases might still be present and lead to idling errors.
The logical $\ket{00}$ state is thus  the ground state of the joint Hamiltonian, fixing the sign such that the overlap with the $\ket{000}$ simulation state is positive, where the three states in $\ket{000}$ correspond to qubit~1, the coupler, and qubit~2, respectively. Similarly, the logical $\ket{11}$ state is the sixth excited state, fixing the sign such that the overlap with the $\ket{101}$ simulation state is positive.
The next step, which is to choose the logical $\ket{01}$ and $\ket{10}$ states, requires greater caution. By construction, these states are degenerate, which means that the eigenstates obtained via a numerical diagonalization are not well-defined. To remove the uncertainty in the eigenstates' definition, we apply L\"{o}wdin's symmetric orthogonalization ~\cite{LowdinOrthogonalization}, which provides an orthonormal set of states with the shortest distance from the original bare states. Similarly to the previous steps, we choose the signs of the computational $\ket{01}$ and $\ket{10}$ states to obtain a positive overlap with the corresponding bare states. 

In summary, we have developed an accurate and efficient method for simulating the evolution of a tunable-coupler system driven by SFQ pulses. The evolution propagator is obtained by simulating the Schr\"{o}dinger evolution under the entire joint Hamiltonian, built out of $n$-level qubits and coupler Hamiltonians, and then projecting it onto the logical subspace, as defined above. We employ the fourth-order Suzuki--Trotter approximation to numerically evaluate the propagator, which results in a variational quantum circuit that we then optimize. In the next section, we present our approach to solving this optimization problem, aimed at delivering high-fidelity CZ and CNOT gates.

\section{Optimization Method}\label{sec:method}

The goal of our SFQ-control optimization is to maximize the fidelity between the generated unitary evolution and the target two-qubit gate. We therefore aim to minimize a cost function $C(\Vec{\theta}) = 1 - F(\Vec{\theta})$, where $F$ is the average gate fidelity and $\Vec{\theta}$ is a set of variational control parameters. More specifically, $\Vec{\theta}$ contains a binary vector with information about the presence or absence of an SFQ pulse at each clock period (and for each qubit), as well as the start and end times of the coupler excursions (i.e., the turning on and off of the coupler), which need to be aligned with the SFQ clock. 

\subsection{Gradient-Based Optimization}

The variational parameters of the introduced cost functions are discrete and, therefore, a straightforward application of gradient-based optimization is not feasible. To deal with this problem, we relax the restriction of the arguments having to be discrete and allow the SFQ pulse amplitudes to take continuous values between 0 and 1. We also let the on and off times of the coupler to take arbitrary values. By permitting the variational parameters to assume continuous values, we are able to compute the gradient of the cost function, but with the undesired result of obtaining nonphysical solutions. To address this issue, we introduce extra penalty terms into the cost function:
\begin{equation}\label{eq:cost}
    C(\Vec{\theta}) = 1 - F(\Vec{\theta}) + P(\Vec{\theta}) + \Phi(\Vec{\theta}).
\end{equation}
Here, 
\begin{equation}
    P(\Vec{\theta}) = \gamma \left[\sum_i \theta_{\mathrm{sfq}}^i(1-\theta_{\mathrm{sfq}}^i) - \sum_j \cos \frac{2\pi \theta_{\mathrm{c}}^j}{T}\right]
\end{equation}
forces the SFQ pulse amplitudes $\theta_{\mathrm{sfq}}^i$ towards physically valid 0 or 1 values, and the coupler start and end times $\theta_{\mathrm{c}}^j$ to align with the ticks of the SFQ clock. The hyperparameter $\gamma$ adjusts the weight of this term, and $T$ is the SFQ-clock period. The last term in the cost function \eqref{eq:cost},
\begin{equation}
    \Phi(\Vec{\theta}) = -\mu \sum_i  \left[ \ln \theta_{\mathrm{sfq}}^i + \ln (1-\theta_{\mathrm{sfq}}^i)\right],
\end{equation}
is a logarithmic smoothing that helps to avoid local minima at the boundaries, where $\mu$ is a smoothing hyperparameter~\cite{murray2010algorithm}. Our empirical investigations have shown that incorporating the smoothing term in addition to regularization provides an advantage in terms of finding the optimal solution faster.

The hyperparameters $\gamma$ and $\mu$ are set to follow a schedule throughout the optimizations. We find that an exponentially increasing (decreasing) schedule $\gamma$ ($\mu$) works well in practice. For the optimizations, we have implemented the following schedule: the initial values are $\gamma=10^{-5}$ and $\mu=1$; we then increase (decrease) the values of $\gamma$ ($\mu$) by a factor of 1.1 after every sequence of 20 parameter updates. We repeat this cycle 150 times for a total of 3000 gradient descent steps (i.e., $150\times 20$).

The optimization processes involve a significant number of parameters. For example, setting the duration to 80~ns and the clock frequency to 40~GHz, we obtain $40 \times 80 = 3200$ amplitudes per qubit. Adding the start and end times of the two coupler excursions brings the total to 6404 variational parameters. To evaluate the gradient efficiently, we have implemented our simulator in JAX~\cite{jax2018github} so as to take advantage of its efficient auto-differentiation. This method allows for the calculation of all partial derivatives in the gradient with a computational cost that is double that of a forward pass that resolves the Schr\"{o}dinger system's evolution. Such an approach is more efficient than the traditional approach of using genetic algorithms~\cite{liebermann2016optimal, jokar2021practical}. As an optimizer, we employ the second-order L-BFGS-B algorithm implemented in SciPy~\cite{2020SciPy-NMeth}, as it effectively scales to the number of variational parameters in the optimizations, automatically tunes the learning rate, and permits the restriction of variational parameter ranges.

We perform a random search to optimize the hyperparameters that are present. Aside from $\gamma$ and $\mu$, the hyperparameters encompass the total duration of the evolution, the SFQ-clock frequencies and kick angles, the varying counts of steps necessary to turn the coupler on and off, and the number of coupler excursions. We also explore various approaches for randomized initialization of the variational parameters.

\subsection{Analytical Decomposition}

Instead of conducting optimization directly, one can use an alternative method to build CZ and CNOT gates out of simpler native gates. This ``modular'' approach allows us to generate the target gates using less memory to store the SFQ pulses~\cite{mukhanov1993rapid}. 

The native two-qubit interaction of the tunable-coupler architecture is well-described by a Hamiltonian of the form
\begin{equation}\label{fsim_hamiltonian}
    H_{\widetilde{\rm{fSim}}} = c_{\textrm{iswap}}(XX+YY)/2 + c_{\textrm{z}}(ZI+IZ) + c_{\textrm{zz}}ZZ ,
\end{equation}
which generates an fSim-like gate. We note that the $(ZI+IZ)$ term commutes with the iSWAP and the $ZZ$ terms, and can therefore be factored out and compensated for in a decomposition.

It has been shown~\cite{GoogleSupremacy} that a CZ gate (and therefore a CNOT gate) can be decomposed into two layers of fSim gates situated between three layers of single-qubit gates. For instance, the circuit

\begin{figure}[tbp]
    \centering
    \includegraphics[width=\columnwidth]{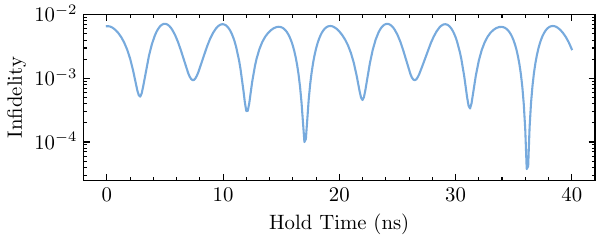}
    \caption{Infidelity of the fSim gate as a function of the hold time. The number of steps for the on and off ramps is 64, each with a duration of 0.05~ns. The total fSim gate duration is therefore equal to the hold time plus 6.4~ns.} 
    \label{fig:fsim}
\end{figure}

\begin{figure}[hb]
    \centering
    \includegraphics[width=\columnwidth]{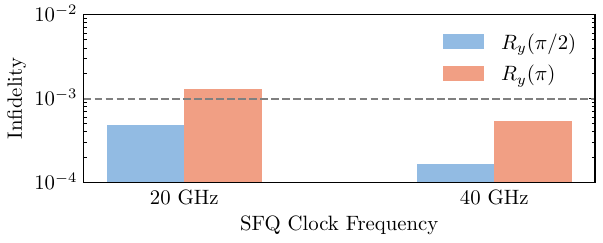}
    \caption{Infidelities of single-qubit rotations for different clock frequencies and rotation angles.} 
    \label{fig:ry_results}
\end{figure}

\vspace{10pt}
\begin{adjustbox}{width=1.0\columnwidth}
\begin{quantikz}
\gate{R_x(\xi)} & \gate[2]{\Gamma (\theta,\phi)}  & \gate{R_x(2\alpha)} & \gate[2]{\Gamma (-\theta,\phi)} & \gate{R_x(\xi)} & \\
\gate{R_x(\eta)} & & \qw & & \gate{R_x(-\eta)} & \\
\end{quantikz}
\end{adjustbox}
produces a CZ gate up to identical (i.e., being the same on both qubits) single-qubit $Z$ rotations. Here, 
\begin{equation}
   \Gamma(\theta, \phi) = 
   e^{-i\theta(XX+YY)/2}e^{-i\phi ZZ/4}
\end{equation}
is essentially an fSim gate with compensated single-qubit $Z$ rotations, and
\begin{subequations}
\begin{align}
    &\xi = \arctan(\frac{\tan\alpha\cos\theta}{\cos\phi/2}) +\frac{\pi}{2}[1-\mathrm{sgn}(\cos\phi/2)] ,\\
    &\eta = \arctan(\frac{\tan\alpha\sin\theta}{\sin\phi/2}) +\frac{\pi}{2}[1-\mathrm{sgn}(\sin\phi/2)] ,\\
    &\alpha = \arcsin\sqrt{\frac{1/2-\sin(\phi/2)^2}{\sin(\theta)^2-\sin(\phi/2)^2}} .
\end{align}
\end{subequations}
Furthermore, the decomposition described is not unique; instead, it represents a continuous family of decompositions, each of which is valid if either of the following conditions is satisfied:
\begin{subequations}
    \begin{align}
        |\sin\theta| \leq \sin\pi/4 \leq |\sin\phi/2| ,\\
        |\sin\phi/2| \leq \sin\pi/4 \leq |\sin\theta|.
    \end{align}
\end{subequations}
Therefore, we have valid decompositions that correspond to different values of the iSWAP and $ZZ$ angles in the fSim Hamiltonian~\eqref{fsim_hamiltonian}. 
We leverage this degree of freedom in the decomposition to produce CZ or CNOT gates of the highest fidelity. To this end,  we study the average gate error of the fSim gate as a function of the hold time between the on and off ramps. As shown in \cref{fig:fsim}, the error oscillates, due to the coherent nature of leakage. We have chosen for our decomposition a hold time of 17~ns, which corresponds to an fSim gate duration of 23.4~ns and a fidelity approximately equal to 0.9999. Furthermore, we note that this value of the hold time corresponds to an iSWAP rotation angle close to $\pi/4$, which produces approximately an $\sqrt{\textrm{iSWAP}}$ gate, which is the shortest fSim-type gate that can generate the desired decomposition, given the small value of $\phi$ for the native two-qubit interaction. 
Lastly, to produce high-fidelity single-qubit rotations required in the decomposition, we rely on the recently developed~\cite{shillito2023compact} DRAG-inspired, single-qubit, SFQ-control method, which has been shown to deliver $X$ and $Y$ rotations with average gate fidelities above 0.9999 for a single-qubit transmon architecture. In the two-qubit tunable-coupler architecture, due to the higher dimension of the Hilbert space, this method generates single-qubit gates with fidelities greater than 0.999, as illustrated in \cref{fig:ry_results}.

In the next section, we provide a summary of the results obtained from gradient-based and analytical decomposition methods and discuss their respective advantages and disadvantages.

\section{Results}\label{sec:results}

\begin{figure}[!tbp]
    \centering
   \includegraphics[width=\columnwidth]{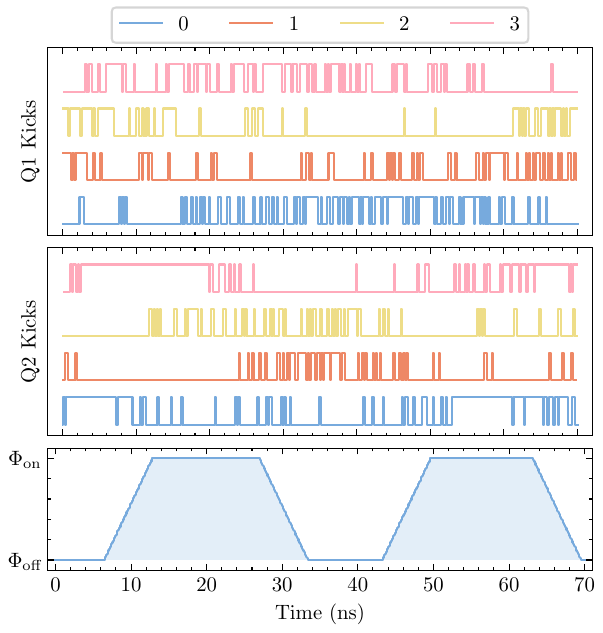}     
    \caption{Control sequences used to produce a CZ gate using gradient-based optimization. Single flux quantum kick sequences on the first qubit (top panel) and the second qubit (middle panel), as well as coupler excursion (bottom panel) between the off point and the on point. The value of the kick angle is $\pi/100$, the SFQ-clock frequency is 20~GHz, and the coupler's on and off flux values are $\Phi_{\textrm{off}}=0.352~\Phi_0$ and $\Phi_{\textrm{on}}=0.376~\Phi_0$, respectively. The four kick plots in the top and middle panels correspond to the four possible SFQ-clock slots during a qubit period. Each step in a single SFQ sequence corresponds to a sequence of kicks repeated at each qubit rotation. } 
    \label{fig:cnot_sequence}
\end{figure}

\begin{figure}[!tbp]
    \centering
    \includegraphics[width=\columnwidth]{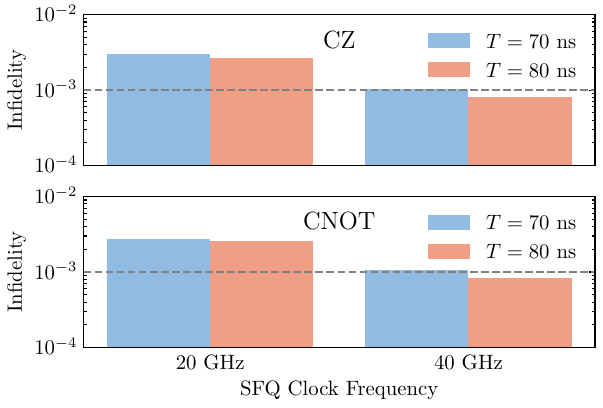}
    \caption{Infidelities of two-qubit gates for different clock frequencies and durations.} 
    \label{fig:cz_cnot_results}
\end{figure}

Using the gradient-based optimization approach described in the previous section, we obtain CZ and CNOT gates with average gate fidelities greater than 0.999.

\Cref{fig:cnot_sequence} shows an example of optimized  SFQ-control sequences for a CZ target gate with a duration of 70~ns. Single flux quantum kicks are delivered to the two qubits (labelled ``Q1'' and ``Q2''), and two coupler excursions comprise the two-qubit interaction. We note that, although the coupler's on and off ramps look linear in the figure, they are in fact piecewise-constant, generated by the addition or removal of flux quanta at every clock period. Furthermore, since the SFQ-clock frequency is four times larger than the qubit frequency, we plot four separate curves for each qubit drive, corresponding to the four possible slots where kicks may happen. This results in better interpretability of the pattern of kick sequences and the memory needed to store the sequences.

In \cref{fig:cz_cnot_results}, we show the infidelities of our optimized sequences for the target CZ and CNOT gates. We use clock frequencies of 20 and 40~GHz, which correspond to 4$\times$ and 8$\times$ the qubit frequency. The SFQ-kick angle for the 20~GHz frequency is $\pi/100$, while for 40~GHz it is $\pi/200$. The results are presented for two gate durations.
The figure shows that higher SFQ frequencies and higher durations generate better results, as expected.

\begin{figure}[b!]
    \centering
    \includegraphics[width=\columnwidth]{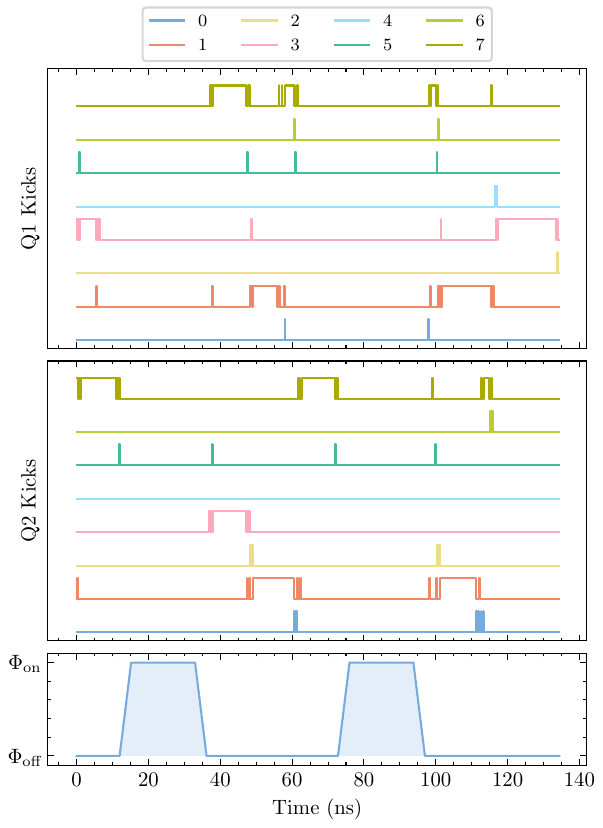}
    \caption{Control sequences used to produce a CZ gate using the decomposition method. Single flux quantum kick sequences on qubit~1 (top panel) and qubit~2 (middle panel), as well as coupler excursion (bottom panel) between the off point and the on point. The value of the kick angle is $\pi/100$, the SFQ-clock frequency is 40~GHz, and the coupler's on and off flux values are $\Phi_{\textrm{off}}=0.352~\Phi_0$ and $\Phi_{\textrm{on}}=0.376~\Phi_0$, respectively. The eight kick plots in the top and middle panels correspond to the eight possible SFQ-clock slots during a qubit period. Each step in a single SFQ sequence corresponds to a sequence of kicks repeated at each qubit rotation.} 
    \label{fig:cz_decomposed_sequence}
\end{figure}

Finally, we apply the analytical decomposition method to produce CZ and CNOT gates. Our resulting gate constructions yield average gate fidelities close to 0.999. An example of such a sequence for a CZ gate is shown in \cref{fig:cz_decomposed_sequence}. As illustrated in the figure, despite the longer duration, these sequences clearly have a very compact representation and can be stored using less memory than the ones obtained by numerical optimization. For comparison, a\, $T=80$~ns optimized sequence at 40~GHz requires 3200 bits per qubit. In contrast, using the encoding described in Ref.~\cite{shillito2023compact}, the number of bits required to store the sequence in \cref{fig:cz_decomposed_sequence} can be estimated to be less than 200 bits per qubit. For the coupler, both methods can utilize an efficient binary encoding of the start and end times of the two-qubit interactions.

\section{Conclusion}\label{sec:conclusion}

In this work we have presented a method to build two-qubit gates, such as fSim, CZ, and CNOT gates, using a tunable-coupler architecture and single flux quantum control.
We examined two approaches: (i) gradient-based optimization of the quantum channel produced by the SFQ sequences, using continuous embedding of the discrete control variables; and (ii) analytical decomposition of a CZ (or CNOT) gate into a pair of fSim gates and single-qubit rotations. We achieved average gate fidelities on the order of 0.9999 for the fSim gate and 0.999 for the CZ and CNOT gates, which are comparable to state-of-the-art results reported for microwave-control gates. Furthermore, the demonstrated gate durations are less than 80~ns for the gradient-based approach and close to 130~ns when using the decomposition method, and could be reduced even further if higher SFQ-clock frequencies are used. This makes our approach a promising and scalable alternative to achieving FTQC.
The idling point of our system is constructed such that the qubits have the same target frequency, which makes the realization of single-qubit gates straightforward, without the need for qubit-specific optimization. We attained single-qubit gate fidelities between 0.999 and 0.9999.
The tunable-coupler system, with qubits idling at an equal frequency, presents a promising approach for the scaling of quantum processors to a larger number of qubits. Future investigation should focus on extending this method to accommodate even larger quantum systems.

\section*{ACKNOWLEDGEMENTS} We thank our editor, Marko Bucyk, for his careful review and editing of the manuscript. P.~R. acknowledges the financial support of Mike and Ophelia Lazaridis, Innovation, Science and Economic Development Canada (ISED), and the Perimeter Institute for Theoretical Physics. Research at the Perimeter Institute is supported in part by the Government of Canada through ISED and by the Province of Ontario through the Ministry of Colleges and Universities.




\begin{thebibliography}{99}

\bibitem{krinner2019engineering} S. Krinner, S. Storz, P. Kurpiers, P. Magnard, J. Heinsoo, R. Keller, J. Luetolf, C. Eichler, and A. Wallraff, 
Engineering cryogenic setups for 100-qubit scale superconducting circuit systems, EPJ Quantum Technology \textbf{6}, 2 (2019).

\bibitem{lin1995timing}
J.-C. Lin and V. Semenov, Timing circuits for rsfq digital systems, IEEE transactions on applied superconductivity \textbf{5}, 3472 (1995).

\bibitem{mancini1999phase}
C. A. Mancini and M. F. Bocko, 
Phase-locked operation of rsfq ring oscillators, 
Superconductor Science and Technology \textbf{12}, 789 (1999).

\bibitem{leonard2019digital}
E. Leonard Jr, M. A. Beck, J. Nelson, B. G. Christensen,
T. Thorbeck, C. Howington, A. Opremcak, I. V. Pechenezhskiy,
K. Dodge, N. P. Dupuis, \emph{et al.},
Digital coherent control of a superconducting qubit, 
Physical Review Applied \textbf{11}, 014009 (2019).

\bibitem{li2019hardware}
K. Li, R. McDermott, and M. G. Vavilov, 
Hardware-efficient qubit control with single-flux-quantum pulse sequences, 
Physical Review Applied \textbf{12}, 014044 (2019).

\bibitem{mcdermott2014accurate}
R. McDermott and M. G. Vavilov, 
Accurate qubit control with single flux quantum pulses, 
Physical Review Applied \textbf{2}, 014007 (2014).

\bibitem{howe2022digital}
L. Howe, M. Castellanos-Beltran, A. Sirois, D. Olaya, J. Biesecker, P. Dresselhaus, S. P. Benz, and P. Hopkins, 
Digital control of a superconducting qubit using a Josephson pulse generator at 3K, 
PRX Quantum \textbf{3}, 010350 (2022).

\bibitem{liu2023single}
C.-H. Liu, A. Ballard, D. Olaya, D. R. Schmidt, J. Biesecker, T. Lucas, J. Ullom, S. Patel, O. Rafferty, A. Opremcak, \emph{et al.},
Single flux quantum-based digital control of superconducting qubits in a multichip module, 
PRX Quantum \textbf{4}, 030310 (2023).

\bibitem{liebermann2016optimal}
P. J. Liebermann and F. K. Wilhelm, 
Optimal qubit control using single-flux quantum pulses, 
Physical Review Applied \textbf{6}, 024022 (2016).

\bibitem{mcdermott2018quantum}
R. McDermott, M. Vavilov, B. Plourde, F. Wilhelm, P.~Liebermann, O.~Mukhanov, and T. Ohki, 
Quantum--classical interface based on single flux quantum digital logic, 
Quantum Science and Technology \textbf{3}, 024004 (2018).

\bibitem{dalgaard2020global}
M. Dalgaard, F. Motzoi, J. J. S{\o}rensen, and J. Sherson, 
Global optimization of quantum dynamics with AlphaZero deep exploration, 
NPJ Quantum Information \textbf{6}, 6 (2020).

\bibitem{jokar2021practical}
M. R. Jokar, R. Rines, and F. T. Chong, 
Practical implications of SFQ-based two-qubit gates, 
Proceedings of the 2021 IEEE International Conference on Quantum Computing and Engineering (QCE), pp. 402--412 (2021).

\bibitem{jokar2022digiq}
M. R. Jokar, R. Rines, G. Pasandi, H. Cong, A. Holmes, Y. Shi, M. Pedram, and F. T. Chong, 
DigiQ: A scalable digital controller for quantum computers using SFQ logic, 
Proceedings of the 2022 IEEE International Symposium on High-Performance Computer Architecture (HPCA), pp. 400--414 (2022).

\bibitem{wang2023single}
Y. Wang, W. Gao, K. Liu, B. Ji, Z. Wang, and Z. Lin, 
Single-flux-quantum-activated controlled-Z gate for transmon qubits, 
Physical Review Applied \textbf{19}, 044031 (2023).

\bibitem{Yan2018}
F. Yan, P. Krantz, Y. Sung, M. Kjaergaard, D. L. Campbell, T. P. Orlando, S. Gustavsson, and W. D. Oliver, 
Tunable coupling scheme for implementing high-fidelity two-qubit gates, 
Phys. Rev. Appl. \textbf{10}, 054062 (2018).

\bibitem{GoogleSupremacy}
F. Arute, K. Arya, R. Babbush, D. Bacon, J. C. Bardin, R. Barends, R. Biswas, S. Boixo, F. G. Brandao, D. A. Buell, \emph{et al.}, 
Quantum supremacy using a programmable superconducting processor,
Nature \textbf{574}, 505 (2019).

\bibitem{google2023suppressing}
R. Acharya, I. Aleiner, R. Allen, T. I. Andersen, M. Ansmann, F. Arute, K. Arya, A. Asfaw, R. B. J. Atalaya, \emph{et al.}, 
Suppressing quantum errors by scaling a surface code logical qubit,
Nature \textbf{614}, 676 (2023).

\bibitem{stehlik2021tunable}
J. Stehlik, D. M. Zajac, D. L. Underwood, T. Phung, J. Blair, S. Carnevale, D. Klaus, G. Keefe, A. Carniol, M. Kumph, \emph{et al.}, 
Tunable coupling architecture for fixed-frequency transmon superconducting qubits,
Physical Review Letters \textbf{127}, 080505 (2021).

\bibitem{sete2024error}
E. A. Sete, V. Tripathi, J. A. Valery, D. Lidar, and J. Y. Mutus, 
Error budget of a parametric resonance entangling gate with a tunable coupler, 
Physical Review Applied \textbf{22}, 014059 (2024).

\bibitem{sung2021realization}
Y. Sung, L. Ding, J. Braumüller, A. Vepsäläinen, B. Kannan, M. Kjaergaard, A. Greene, G. O. Samach, C. McNally, D. Kim, \emph{et al.}, Physical Review X \textbf{11}, 021058 (2021).

\bibitem{shillito2023compact}
R. Shillito, F. Hopfmueller, B. Kulchytskyy, and P. Ronagh, 
Compact pulse schedules for high-fidelity single-flux quantum qubit control,
arXiv:2309.04606 (2023).

\bibitem{kirichenko2022system}
A. F. Kirichenko, A. Jafari-Salim, P. Truitt, N. K. Katam, C. Jordan, and O. A. Mukhanov, 
System and method of flux bias for superconducting quantum circuits,
U.S. Patent Application 17/838,207 (2022).

\bibitem{vool2017introduction}
U. Vool and M. Devoret, 
Introduction to quantum electromagnetic circuits,
Int. J. Circuit Theory Appl. \textbf{45}, 897 (2017).

\bibitem{LowdinOrthogonalization}
I. Mayer,
On L\"{o}wdin’s method of symmetric orthogonalization,
Int. J. Quantum Chem. \textbf{90}, 63 (2002).

\bibitem{murray2010algorithm}
W. Murray and K.-M. Ng,
An algorithm for nonlinear optimization problems with binary variables,
Comput. Optim. Appl. \textbf{47}, 257 (2010).

\bibitem{jax2018github}
J. Bradbury, R. Frostig, P. Hawkins, M. J. Johnson, C. Leary, D. Maclaurin, G. Necula, A. Paszke, J. VanderPlas, S. Wanderman-Milne, and Q. Zhang, JAX: composable transformations of Python+NumPy programs (2018).

\bibitem{2020SciPy-NMeth}
P. Virtanen, R. Gommers, T. E. Oliphant, M. Haberland, T. Reddy, D. Cournapeau, E. Burovski, P. Peterson, W. Weckesser, J. Bright, S. J. van der Walt, M. Brett, J. Wilson, K. J. Millman, N. Mayorov, A. R. J. Nelson, E. Jones, R. Kern, E. Larson, C. J. Carey, ˙I. Polat, Y. Feng, E. W. Moore, J. VanderPlas, D. Laxalde, J. Perktold, R. Cimrman, I. Henriksen, E. A. Quintero, C. R. Harris, A. M. Archibald, A. H. Ribeiro, F. Pedregosa, P. van Mulbregt, and SciPy 1.0 Contributors,
SciPy 1.0: Fundamental Algorithms for Scientific Computing
in Python,, Nature Methods \textbf{17}, 261 (2020).

\bibitem{mukhanov1993rapid}
O. A. Mukhanov, 
Rapid single flux quantum (rsfq) shift register family,
IEEE Trans. Appl. Supercond. \textbf{3}, 2578 (1993).


\end{thebibliography}
\end{document}